\documentstyle[aps,prd,preprint,tighten,mathrsfs,latexsym,amsmath,amssymb,
amstext,amsthm,amsfonts,epic,eepic,epsfig,axodraw,graphicx]{revtex}

\newcommand{\sslash}[1]{{#1}\!\!\!\slash}

\newcommand{\id}{{\hbox{{\rm 1}\kern-.24em\hbox{\rm l}}}}
\newcommand{\one}{{\hbox{{\rm 1}\kern-.24em\hbox{\rm l}}}}
\newcommand{\uni}{{\hbox{{\rm 1}\kern-.24em\hbox{\rm l}}}}
\newcommand{\pslash}{{\hbox{{\rm p}\kern-.5em\hbox{\slash}}}}
\newcommand{\Aslash}{{\hbox{{\rm A}\kern-.5em\hbox{\slash}}}}
\newcommand{\partialslash}{{\hbox{{$\partial$}\kern-.5em\hbox{\slash}}}}

\begin{document}
\draft

\title{Quantum Field Theory Treatment of Neutrino Oscillations in Vacuum and in Matter}

\author{Diego Pallin\thanks{E-mail: {\tt diego@theophys.kth.se}} and
 H{\aa}kan Snellman\thanks{E-mail: {\tt snell@theophys.kth.se}}}
\address{Division of Mathematical Physics, \\ Department of
Physics, Royal Institute of Technology, KTH-SCFAB,\\ SE-10961 Stockholm, Sweden}
\date{\today}

\maketitle

\begin{abstract}
We study neutrino oscillations in vacuum and in matter using field theoretical 
methods and wave-packets. In particular, we calculate the neutrino propagator in
the presence of matter with constant density for the case of two flavors, obtaining 
the resonance formula. In the extreme relativistic limit, the result of the usual 
quantum mechanical treatment is recovered with interesting, but small modifications. 
\end{abstract}

\section{Introduction}
\label{sec:intr}

Evidence for neutrinos having small, but finite, masses is accumulating fast.
The main paradigm for this is the idea, first proposed by Pontecorvo
\cite{bib:pontecorvo}, that if the neutrino flavors were a
superposition of massive eigenstates, then 
the neutrinos could oscillate between these different flavors. There are many 
experiments looking for neutrino oscillations: solar 
neutrino measurements, atmospheric neutrino measurements, reactor 
experiments, and accelerator experiments.  Several of them have claimed 
evidence for neutrino oscillations.  The 
final breakthrough came in June 1998 when the Super-Kamiokande 
collaboration reported strong evidence for neutrino oscillations from the 
atmospheric neutrino {\small UP-DOWN} asymmetry.  
The measurements on the depletion of atmospheric muon neutrino flux fit 
well to a two flavor $\nu_\mu\leftrightarrow \nu_\tau$ oscillation model. This 
summer the SNO collaboration \cite{bib:SNO} showed that the solar neutrino deficit, 
pioneered by Davies \cite{bib:davies}, was due 
to conversion of electron neutrinos to mu- and tau neutrinos.  Recently, also
the KamLAND experiment \cite{bib:kamland} shows evidence for neutrino oscillations.

In the standard quantum mechanical treatment of neutrino oscillations, the mass eigenstates 
are assumed to be relativistic and to have the same momentum, and thus, 
different energies.  The familiar quantum mechanical model describing the flavor mixing 
process has several conceptual difficulties compared to quantum field theory models, see 
Refs. \cite{bib:Beuthe,bib:Cardall1,bib:PhysRevD48p4310,bib:Grimus,bib:Rich}.  
For example, energy momentum conservation in the production and detection 
processes that forces neutrinos to be in a mass eigenstate is incompatible 
with neutrino oscillations.  The neutrino oscillation probability is 
independent of the details concerning the production and detection 
processes only in the extremely relativistic limit.  Hence, in the case 
that some of the mass eigenstates cannot be considered to be extremely 
relativistic, one has to use quantum field theory.  Of course, the quantum field theory expression must 
reproduce the quantum mechanics oscillation probability in the ultra-relativistic 
limit.  A very detailed review regarding different aspects and questions of 
neutrino oscillations in quantum field theory can be found in Ref. \cite{bib:Beuthe}.  Another 
interesting question which is important in both a quantum mechanical or quantum field theory treatment is if 
there exists a Fock space for the flavor eigenstates, since there exists a 
Fock space for the mass eigenstates.  This question has been 
discussed by Giunti {\it et al.\/} \cite{bib:PhysRevD45p2414}.  Fuji {\it et al.\/} \cite{bib:Fujii} 
give arguments that a Fock space of flavor 
neutrinos does not exist.

When neutrinos propagate through matter their behavior may be affected 
significantly, as was pointed out by Wolfenstein \cite{bib:Wolfenstein} in 
1978 and emphasized by Mikheyev and Smirnov \cite{bib:msw} as being 
important for neutrino oscillations.  This is due to the fact that in the
presence of matter the effective mass induced by the forward scattering of 
neutrinos by the background changes the flavor oscillating parameters.  A 
quantum mechanical treatment of neutrinos interacting with matter can be 
found in almost all books treating neutrinos.  Peltoniemi and Sipil\"ainen 
\cite{bib:Peltoniemi} have studied neutrino propagation in matter using a 
wave packet approach.  Cardall and Chung \cite{bib:Cardall2} treats 
neutrino oscillations in a static uniform background by quantum field 
theory and show that they recover the quantum mechanical oscillation 
amplitude in the relativistic limit.  Also Fuji {\it et al.\/} \cite{bib:Fujii} treats 
neutrino oscillations in a static matter background using Bogoliubov 
transformations.

In the present paper, we give a short presentation of the use of field 
theory methods to describe neutrino oscillations in vacuum and in matter.  
For neutrino oscillations in matter we carry out the calculations for the 
case of two flavors in detail, which displays the main features of our 
approach.  Our aim is in particular to calculate explicitly the singularity structure 
of the neutrino Green's function in the presence of matter to study the 
resonace formula.

In Sec.~\ref{sec:form}, we review the basic formalism to be used.  
Section~\ref{sec:flavor} contains an elaboration of the use of wave-packets 
for flavor states.  We then calculate the oscillation amplitude for the 
case of vacuum oscillations in Sec.~\ref{sec:vacuum}, and then, treat 
explicitly the case of two flavors in the presence of matter with constant 
density in Sec.~\ref{sec:matter}.

\section{Basic Formalism}
\label{sec:form}

The equation of motion for the neutrino field $\nu_j$, with mass $m_{j}$, can be written in the 
following form
\begin{equation}
( i \sslash{\partial} - m_{j})\nu_{j}(x) = \chi_{j}(x),
	\label{matter01}
\end{equation}
where we have introduced the source term $\chi_j(x)$, which can be obtained 
from the complete Lagrangian density of the standard electroweak theory in 
the unitary gauge.  The interaction part of primary interest to us is given 
by

\begin{eqnarray}
\label{eq:LagrangianDensityInteractionFlavorBasis}
{\cal L}_{\rm{int}} (x) &=&\sum_{\beta=e,\mu,\tau}\left[-\frac{g}{\sqrt{2}}\overline{\nu}_{\beta L}
(x)\gamma^\alpha \psi_{\beta L}(x)W_\alpha(x) -\frac{g}{\sqrt{2}}\overline{\psi}_{\beta L}(x)\gamma^\alpha\nu_{\beta L}(x)W^\dagger_\alpha(x) \right.-\nonumber \\
& &\left.\frac{g}{2\cos\theta_W}\overline{\nu}_{\beta L}(x)\gamma^\alpha\nu_{\beta L}(x)Z_\alpha(x) \right]
-\sum_{j=1}^{3} \frac{1}{v}m_{\nu_j}\overline{\nu}_{j}(x)\nu_{j}(x)\sigma,
\end{eqnarray}
where $g$ is a dimensionless coupling constant and $W_\alpha(x)$ and $Z_{\alpha}$ are the fields that describes 
the $W$ and $Z$ bosons with spin 1. The last part contains the Higgs field $\sigma$ and the electroweak vacuum expectation value $v$. We observe that the interaction part contains the auxiliary flavor  
fields $\nu_{\beta}$. These fields are expressed in terms of the mass eigenfields $\nu_j$ by the 
relations $\nu_{\beta}(x)=\sum_jU_{\beta j}\nu_j(x)$ and 
($\overline{\nu}_{\beta}(x)=\sum_jU^\ast_{\beta j}\overline{\nu}_j(x)$), where $U$ is unitary 
leptonic mixing matrix to be specified below.  Originally the $U$-matrix is defined 
only for the left handed fields.  For simplicity we will assume that the 
same transformation also applies for the right handed fields.  Inserting 
the expressions for $\nu_{\beta}$ and $\overline{\nu}_{\beta}$ expressed in 
terms of the mass eigenfields into the interaction Lagrangian, one obtains

\begin{eqnarray}
\label{eq:LagrangianDensityInteractionMassBasis}
{\mathcal L}_{\textrm{int}} (x) &=&\sum_{\beta=e,\mu,\tau}\sum_{j=1}^3 \Big[ -\frac{g}{\sqrt{2}}\left[ U^\ast_{\beta j}
\overline{\nu}_{j L}(x)\gamma^\alpha \psi_{\beta L}(x)W_\alpha(x)\right.+ \nonumber \\
& &\left.\overline{\psi}_{\beta L}(x)\gamma^\alpha U_{\beta j}\nu_{j L}(x)W^\dagger_\alpha(x) \right] \Big] - 
\frac{g}{2\cos\theta_W}\sum_{k=1}^3\overline{\nu}_{k L}(x)\gamma^\alpha \nu_{k L}(x)Z_\alpha(x).
\end{eqnarray}
We have omitted the part of the interaction Lagrangian containing the $\sigma$ field in the Higgs' mechanism. This 
Lagrangian gives 
\begin{eqnarray}
\label{eq:sourceterm}
\chi_j(x)&=&\sum_{\beta=e,\mu,\tau}\left[\frac{g}{\sqrt{2}}U^\ast_{\beta 
j}\gamma^\alpha \psi_{\beta L}(x)W_\alpha(x) \right] 
+\frac{g}{2\cos\theta_W}\gamma^\alpha \nu_{j L}(x)Z_\alpha(x) ,  
\end{eqnarray} 
describing the fundamental processes that go on in the source and the 
detector.  The first two terms in Eq.~\ref{eq:LagrangianDensityInteractionMassBasis}
 define the flavor combinations through the 
intermediate boson interaction containing the $U$ matrix components.  This 
mixture in the interaction describes that the weak bosons create a coherent
superposition of neutrino mass eigenfields at the vertex. At energy scales where
the neutrino rest masses are small compared to their total energy, this can give
rise to neturino oscillations. We also observe
that the different neutrino fields $\nu_j(x)$ are coupled to each other via 
the neutral current interaction.  At the energy scale we are going to 
consider, the amplitudes describing the detection processes, e.g.  in the 
SNO detector, only appear with the second order terms in the coupling 
constants.  These processes comprise the elastic scattering of neutrinos on 
electrons, the inverse beta decay and the deuteron break up.  To describe 
these processes we use the effective second order Lagrangian, obtained by 
neglecting the momentum of the intermediate bosons.  This means that we 
insert the weak hadronic currents coupled to the neutrinos with the Fermi 
coupling constant $G_{F}$, and also use the analogue of this for the purely 
leptonic scattering in the Cherenkov detectors.  Finally, we also use this 
effective Lagrangian to describe neutrino interactions with matter, neglecting recoil effects.
The effective Lagrangian for neutrinos interacting with matter is

\begin{equation}
{\cal L}(x)^{{\rm matter}}_{\rm eff} = - \sqrt{2}G_{F}\sum_{j=1}^{3}U_{ej}^{*} \bar \nu_{jL}(x) \gamma^{0}\nu_{jL}(x) 
U_{ej}N_{e}(x) 
+ \frac{1}{\sqrt{2}}G_{F}\sum_{j=1}^{3}\bar \nu_{jL}(x)\gamma^{0} \nu_{jL}(x) N_{n}(x),
\label{eq:sourcematter}
\end{equation}
where $N_{e}(x)$ and $N_{n}(x)$ are the electron and neutron density 
operators, respectively.  In our further discussion, we will only consider 
the case when these densities are constant and take the neutrino fields to 
be the renormalized effective fields.

\section{The Flavor States in a Wave Packet Approach} 
\label{sec:flavor}
To zeroth order in the coupling constant $g$, the uncoupled Dirac equations for neutrinos 
$(i\sslash{\partial}-m_j)\nu_j(x)=0$, has the following free field solutions \cite{bib:peskin}: 
\begin{equation} 
\nu_j(x)=\int\frac{d^3p}{(2\pi)^3}\frac{1}{\sqrt{2E_j({\bf{p}})}}\sum_{s}\left(a^s_j({\bf p}) 
u^s_j(p)e^{-ip_{j}x}+b^{s\dagger}_j({\bf{p}})v^s_j(p)e^{ip_{j}x}\right),  
\end{equation} 
where $u_{j}$ and $v_{j}$ are particle and antiparticle solutions to the 
Dirac equation.  The conjugate field $\overline{\nu}_j(x)$ has an analogous 
expansion.  This expression for the neutrino mass field in terms of the 
operators $a^s_j({\bf{p}})$ and $b^{s\dagger}_j({\bf{p}})$ can be inverted 
for the operators in terms of the fields in the usual way.  In our view, 
only the mass eigenstate fields are those that build up the Fock space 
structure.  The auxiliary ``flavor fields'' are defined as mixtures by the 
transformations given above, and are introduced for convenience.  These 
flavor fields, of course, enter formally into the Lagrangian, as they 
represent the mixture produced by the weak interactions.  However, they 
cannot be considered to represent asymptotic fields that are carrying 
irreducible representations of the Poincar{\'e} or Lorentz groups.  

For our purposes we want to formulate the field theory slightly 
differently.  We introduce the notation

\begin{equation}
	a^s_j({\bf p},t) =a^s_j({\bf p}) e^{-iE_{j}({\bf p})t},
	\label{field1}
\end{equation}
where $E_{j}({\bf p})=\sqrt{{\bf{p}}^{2}+m_{j}^{2}}$. The expansion can then be written
\begin{equation} 
\nu_j(x)=\int\frac{d^3p}{(2\pi)^3}\frac{1}{\sqrt{2E_j({\bf{p}})}}\sum_{s}\left(a^s_j({\bf p}, t ) 
u^s_j(p)e^{i{\bf p}\cdot{\bf x}}+b^{s\dagger}_j({\bf{p}},t) v^s_j(p)e^{-i{\bf p}\cdot{\bf x}}\right).  
\end{equation} 
The time dependent operators $a^{s \dagger}_j({\bf p},t)$ can be projected out of the neutrino fields as
\begin{equation}
	a^{s \dagger}_j({\bf p},t) = \frac{1}{\sqrt{2E_{j}({\bf p})}} u^{s}_{j}(p)\int d^{3}x e^{i\bf{p}\cdot \bf{x}}\nu^{\dagger}_{j}(t,\bf{x}).
	\label{field2}
\end{equation}
These operators and their Hermitean conjugates create and annihilate particles of definite 
three-momentum ${\bf p}$, at time $t$. 

A plane wave basis neutrino state $|\nu_j;r,{\bf p},t\rangle$ with mass $m_j$, momentum ${\bf p}$ 
and spin $r$ at time $t$ is obtained by acting with $a^{r\dagger}_j({\bf p},t)$ on the vacuum 
state $|0\rangle$, defined such that $\langle 0|0\rangle=1$. 
In order for $|\nu_j;r,{\bf p},t \rangle$ to be covariantly normalized at equal times, i.e., 
$\langle\nu_j;s,{\bf q},t|\nu_j;r,{\bf p},t\rangle = 2E_j({\bf{p}})(2\pi)^3\delta^{(3)}({\bf p}- {\bf q})\delta^{rs}$, 
the definition for $|\nu_j;r,{\bf{p}}, t \rangle$ is
\begin{equation}
|\nu_j;r,{\bf{p}},t\rangle \equiv \sqrt{2E_j({\bf{p}})}a^{r\dagger}_j({\bf{p}},t)|0\rangle.
\end{equation}

The auxiliary neutrino flavor fields $\nu_\alpha$, where $\alpha=e,\mu,\tau$, that we will 
use to construct our localized states, are expressed in the mass 
eigenfields $\nu_j$, where $j=1,2,3$, by $\nu_\alpha(x) = \sum_{j}U_{\alpha j}\nu_j(x)$. The flavor fields
are constructed with the mass eigenfields having the same three-momentum, ${\bf p}$.
The $U_{\alpha j}$'s are the elements of the unitary mixing matrix $U$, which for three 
generations is given in standard form by

{\small \begin{equation}
U = \left( \begin{array}{ccc} C_2 C_3 & S_3 C_2 & S_2e^{-i\delta} \\ - S_3 C_1 -
S_1 S_2 C_3e^{i\delta} & C_1 C_3 - S_1 S_2 S_3e^{i\delta} & S_1 C_2 \\ S_1 S_3 - S_2
C_1 C_3e^{i\delta} & - S_1 C_3 - S_2 S_3 C_1e^{i\delta} & C_1 C_2 \end{array} \right)
\end{equation}}

\noindent
with $C_i\equiv\cos{\theta_i}$ and $S_i\equiv\sin{\theta_i}$, $i=1,2,3$. 
When the CP-violating phase $\delta=0$, the matrix elements are real and we have
$U^{\ast}_{\alpha j} = U_{\alpha j}$ for $\alpha = e,\mu,\tau$ 
and $j = 1,2,3$. In what follows, we put $\delta=0.$

Let us define $|\nu_{\alpha};r,{\bf{p}},t\rangle \equiv \sum_{j}U^\ast_{\alpha j}|\nu_j;r,{\bf{p}},t)\rangle$. Since we want 
to have an expression for $|\nu_\alpha;r,{\bf{p}},t\rangle$ in terms of $\nu_j^{\dagger}(x)$, we obtain 
\begin{equation}
|\nu_{\alpha};r,{\bf{p}},t\rangle=\sum_{j}U^\ast_{\alpha j}\sqrt{2E_j({{\bf{p}}})}a^{r\dagger}_j({{\bf{p}}},t)|0\rangle 
= \sum_{j}U^\ast_{\alpha j}u^r_j(p)\int d^3x e^{i\bf{p} \cdot \bf{x}} \nu_j^{\dagger}(x)|0\rangle.
\end{equation}

The state $|\nu_{\alpha};r,{\bf{p}},t \rangle$ has a definite momentum ${\bf{p}}$, 
which means that $\Delta{\bf{p}}=\bf{0}$.  According to Heisenberg's 
uncertainty relation, $\Delta p \Delta x \geq \hbar/2$, the state has therefore 
an infinitely wide spread in space.  It also does not have a well-defined 
energy, since $P^{0}$ operating on the state does not satisfy 
$E_{\alpha}=\sqrt{{\bf p}^{2}+m_{\alpha}^{2}}$.  This is of course one
of the causes of the flavor oscillations in this mode of description.

In order to have a Hilbert space state 
that is localized around the momentum ${\bf{P}}$ at position ${\bf{X}}$ and time 
$T$, one has to use a wave-packet with some distribution function 
$F_{\alpha}({\bf{X}},{\bf{P}},{\bf{p}}) = F_{\alpha}({\bf P},{\bf p})e^{-i{\bf X}\cdot {\bf p}}$.  We 
can then write such a state 
$|F_{\alpha}(r,{\bf{P}},{\bf{X}},T)\rangle$, that is localized around 
${\bf{P}}$ at the space time point ${\bf{X}},T$ with spin projection $r$, as
\begin{equation}
|F_{\alpha}(r,{\bf{P}},{\bf{X}},T)\rangle=\int\frac{d^3p}{(2\pi)^3} \frac{F_{\alpha}({\bf{X}},{\bf{P}},{\bf{p}})}{\sqrt{2E_\alpha({\bf{p}})}} 
|\nu_{\alpha};r,{\bf{p}},T\rangle.
\end{equation}
In what follows we will use the subscripts $s$ and $d$ to denote source and detector, respectively.
Inserting our expression for $|\nu_{\alpha};r,{\bf{p}}, T\rangle$, we finally have
\begin{equation}
|F_{\alpha}(r,{\bf{P}_{\ell}},{\bf{X}_{\ell}},T_{\ell})\rangle = 
\int \frac{d^3p}{(2\pi)^3} \frac{F_{\alpha}({\bf{X}_{\ell}},{\bf{P}_{\ell}},{\bf{p}})}
 {\sqrt{2E_\alpha({\bf{p}})}}\sum_{j}U^\ast_{\alpha j}u^r_j(p)\int d^3x e^{i\bf{p} \cdot \bf{x}} 
 \nu_j^{\dagger}(T_{\ell},{\bf x})|0\rangle
\label{state1}
\end{equation}
for $\ell = s,d$. This expression can be taken as the definition of our localized neutrino 
flavor states.  In actual applications, it is convenient to work with 
Gaussian wave packets, with a mean spread $\sigma$ around ${\bf P}$.  
The corresponding wave function in the plane wave basis is
\begin{equation}
   F_{\alpha}({\bf{P}}_{\ell},{\bf{p}}) = \left(\frac{2\pi}{\sigma^{2}_{\ell}}\right)^{2} 
   e^{-\frac{({\bf P}_{\ell}-{\bf p})^{2}}{4\sigma^{2}_{\ell}}} \;\; \ell=  s,d ,
	\label{gaussian1}
\end{equation}
where for simplicity will take the same $\sigma_{\ell}$ for all flavors. 
The states in Eq.~(\ref{state1}) have correct normalization for equal times if we make the interpretation
$E_\alpha({\bf{p}} ) = \sum_jU_{\alpha j}U^\ast_{\alpha j}E_j({\bf{p}})$, i.e., $E_{\alpha}$ is the
expectation value of the Hamiltonian in the state $|\nu_{\alpha};r,{\bf{p}},T\rangle$.

\section{Neutrino Oscillations in vacuum}
\label{sec:vacuum}

The transition amplitude $A^{rs}_{\alpha\beta} \equiv A^{rs}_{\alpha\beta}({\bf{X}}_{s},{\bf{X}}_{d},T_s,T_d, {\bf P}_{s},{\bf P}_{d})$ 
in the rest frame of the detector and the source for a neutrino of 
flavor $\alpha$, momentum around ${\bf P}_{s}$, and spin $r$ to be produced at the source and to be detected as a flavor $\beta$,
 with momentum around ${\bf P}_{d}$, and spin $s$ is given by
\begin{equation}
A^{rs}_{\alpha\beta}=\langle F_{\beta}(s,{\bf{P}}_{d}, {\bf{X}}_{d}, T_d)|F_{\alpha}(r,{\bf{P}}_{s},{\bf{X}}_{s},T_s)\rangle.
\end{equation}
Using the expression for $|F_{\alpha}(r,{\bf{P}_{\ell}},{\bf{X}_{\ell}},T_{\ell})\rangle$ and its conjugate, we have
\begin{eqnarray}
\label{eq:Amplitude}
A^{rs}_{\alpha\beta} &=& \int\frac{d^3 q}{(2\pi)^3}\frac{F_{\beta}^{*}({\bf{X}}_{d},{\bf{P}}_{d},{\bf{q}})}
{\sqrt{2E_\beta({\bf{q}})}}\sum_{j}U_{\beta j}u^{s\dagger}_j(q)\int d^3y e^{-i\bf{q}\cdot \bf{y}} \nonumber\\
& &\times 
\int\frac{d^3 p}{(2\pi)^3}\frac{F_{\alpha}({\bf{X}}_{s},{\bf{P}}_{s},{\bf{p}})}{\sqrt{2E_\alpha({\bf{p}})}}\sum_{k}U^\ast_{\alpha k}u^r_k(p)
\int d^3x e^{i \bf{p} \cdot \bf{x}} \nonumber \\
& &\times 
\langle 0|\nu_j(T_{d},{\bf y})\nu_k^{\dagger}(T_{s},{\bf x})|0\rangle.
\end{eqnarray}
The expression $\langle 0|\nu_j(y)\nu_k^{\dagger}(x)|0\rangle$ is the propagation amplitude for a mass 
eigenstate $k$ being created at $x=(T_{s},\bf{x})$, propagating to $y=(T_{d},\bf{y})$, where a mass eigenstate 
$j$ is being annihilated.
In vacuum, $j$ has to be equal to $k$. On the other hand, when neutrinos propagate 
through matter it is possible that $j \neq k$. Since $T_{d}>T_{s}$ i.e. $y^0>x^0$, we can insert the time-ordering 
operator $T$ obtaining
\begin{eqnarray}
\label{eq:VacuumExpectationValueToTimeOrderedVacuumExpectationValue}
\langle 0|\nu_j(y)\nu_k^{\dagger}(x)|0\rangle \rightarrow\langle 0|T(\nu_j(y)\nu_j^{\dagger}(x))|0\rangle \delta_{jk}.
\end{eqnarray}
The last expression is simply
\begin{eqnarray}
\langle 0|T(\nu_j(y)\nu_j^{\dagger}(x))|0\rangle &=&\langle 0|T(\nu_j(y)\overline{\nu}_j(x))|0\rangle\gamma^0 =
S_j(y-x)\gamma^0,
\end{eqnarray}
where $S_{j}$ denotes the Feynman fermion propagator defined by
\begin{equation}
\label{eq:FeynmanFermionPropagator}
S_j(y-x)=i\int\frac{d^4k}{(2\pi)^4}\frac{(\sslash{k}+m_j)}{k^2-m^2_j+i\epsilon}e^{-ik(y-x)}.
\end{equation}
The integration in the complex $k^0$-plane has to be taken along the whole real axis. 
We can then rewrite the expression for the amplitude as follows
\begin{eqnarray}
\label{eq:AmplitudeIntegrals}
A^{rs}_{\alpha\beta} &=& \sum_{j}U_{\beta j}U^\ast_{\alpha j}
\int\frac{d^3q}{(2\pi)^3}\int\frac{d^3p}{(2\pi)^3}\frac{F_{\beta}^{*}({\bf{X}}_{d}, {\bf{P}}_{d}, 
{\bf{q}})}{\sqrt{2E_\beta({\bf{q}})}} \frac{F_{\alpha}({\bf{X}}_{s}, {\bf{P}}_{s},{\bf{p}})}{\sqrt{2E_\alpha({\bf{p}})}}\nonumber \\
& &\times \int d^3y e^{-i\bf{q} \cdot \bf{y}} \int d^3x e^{i\bf{p}\cdot \bf{x}}\int\frac{d^4k}{(2\pi)^4}e^{-i[k^{0}(T_{d}-T_{s})-\bf{k}\cdot (\bf{y-x})]} \nonumber \\
& &\times i\;\frac{u^{s\dagger}_j(q)(\sslash{k}+m_j)\gamma^0 u^r_j(p)} {k^2-m^2_j+i\epsilon}.
\end{eqnarray}
We next calculate the integrals in the middle in the order ${\bf x},{\bf y}$, and ${\bf k}$. 
This gives
\begin{eqnarray}
\label{eq:ThreeMiddleIntegrals}
& &i\int\frac{dk^{0}}{2\pi} \int\frac{d^3k}{(2\pi)^3} \int d^3y e^{-i\bf{q}\cdot \bf{y}}\int d^3xe^{i\bf{p}\cdot \bf{x}}\frac{(\sslash{k}+m_j)}{k^2-m^2_j+i\epsilon}e^{-ik(y-x)} 
 \nonumber \\
& &=i(2\pi)^3\delta^{(3)}({\bf{p}} - {\bf{q}} ) \int\frac{dk^{0}}{2\pi}e^{-i(T_{d}-T_{s})k^{0}} 
\frac{\left(k^{0}\gamma^{0}-{\bf p}\cdot {\boldsymbol{\gamma}} +m_j\right)}{(k^{0})^{2}-{\bf p}^{2}-m_{j}^{2} +i\epsilon }. \nonumber \\
\end{eqnarray}
Closing the contour in the lower $k^{0}$ half-plane and calculating the residues of the positive energy poles, it follows that
\begin{eqnarray}
\label{eq:AmplitudeIntegrals1}
A^{rs}_{\alpha\beta} &=& \sum_{j}U_{\beta j}U^\ast_{\alpha j}
\int\frac{d^3q}{(2\pi)^3}\int\frac{d^3p}{(2\pi)^3}\frac{F_{\beta}^\ast({\bf{X}}_{d},{\bf{P}}_{d},{\bf{q}})}{\sqrt{2E_\beta({\bf{q}})}} 
\frac{F_{\alpha}({\bf{X}}_{s},{\bf{P}}_{s},{\bf{p}})}{\sqrt{2E_\alpha({\bf{p}})}} \nonumber \\
& &\times 
(2\pi)^3\delta^{(3)}({\bf{p}}-{\bf{q}})e^{-i(T_{d}-T_{s})E_j({\bf{p}})} u^{s\dagger}_j(q) \frac{\left(\sslash{p}+m_j\right)}{2E_j({\bf{p}})} \gamma^0 u^r_j(p).
\end{eqnarray}
We finally obtain
\begin{eqnarray}
\label{eq:AmplitudeIntegrals2}
A^{rs}_{\alpha\beta} &=& \sum_{j}U_{\beta j}U^\ast_{\alpha j}\int \frac{d^3p}{(2\pi)^3}
\frac{F_{\beta}^{*}({\bf{X}}_{d},{\bf{P}}_{d},{\bf{p}})}{\sqrt{2E_\beta({\bf{p}})}}\frac{F_{\alpha}({\bf{X}}_{s},{\bf{P}}_{s},{\bf{p}})}{\sqrt{2E_\alpha({\bf{p}})}} \nonumber \\
& &\times \frac{u^{s\dagger}_j(p) \left(\sslash{p}+m_j\right)\gamma^0 u^r_j(p)}{2E_j({\bf{p}})}e^{-iE_{j}({\bf p})(T_{d}-T_{s})}.
\end{eqnarray}
The part $u^{s\dagger}_j(p) \left(\sslash{p}+m_j\right)\gamma^0 u^r_j(p)$ can be 
simplified by using that $(\sslash{p}-m_j )u^r_j(p)=0$. This gives
\begin{equation}
\label{eq:p_slashPlusM}
u^{s\dagger}_j(p) \left(\sslash{p}+m_j\right)\gamma^0 u^r_j(p) = (2E_j({\bf{p}}))^2\delta^{rs}.
\end{equation}
We finally arrive at the following rather simple expression for the transition amplitude
\begin{equation}
\label{eq:GeneralAmplitude}
A^{rs}_{\alpha\beta} = \int\frac{d^3 p}{(2\pi)^3} 
\frac{F_{\beta}^{*}({\bf{X}}_{d},{\bf{P}}_{d},{\bf{p}})}{\sqrt{2E_\beta({\bf{p}})}}\frac{F_{\alpha}({\bf{X}}_{s},{\bf{P}}_{s},{\bf{p}})}
{\sqrt{2E_\alpha({\bf{p}})}} M^{rs}_{\alpha \beta},
\end{equation}
where
\begin{equation}
	M^{rs}_{\alpha \beta} = \delta^{rs} \sum_{j}U_{\beta j}U^\ast_{\alpha j}2E_j({\bf{p}}) e^{-iE_{j}({\bf p})(T_{d}-T_{s})}.
\end{equation}
The amplitude $A^{rs}_{\alpha\beta}$ in Eq.~(\ref{eq:GeneralAmplitude}) will be 
our standard reference amplitude, when we later calculate the amplitude for neutrino 
oscillations in the presence of matter. Using Gaussian wave packets, it is possible to 
show that we regain the usual neutrino oscillation formula in the ultra-relativistic limit. 
In particular, it is useful to consider the time average of the amplitude, as we do not
know when the neutrinos are emitted. We will not here elaborate further on this point, 
which is treated in Ref. \cite{bib:PhysRevD45p2414}, but continue with the case when neutrinos 
interact with matter.

\section{Neutrinos in matter: the resonance formula}
\label{sec:matter}

\subsection{General discussion}
\label{sec:matter1}
We will next proceed to derive the neutrino oscillation probability for 
neutrinos propagating in matter with constant density.  Although a 
perturbative treatment of neutrino oscillations in matter would give the 
influence of matter for density factors that are small compared to the 
energy differences $\Delta m^{2}/E$ between the neutrinos, the behavior of 
the neutrinos are quite different for densities that are of the same order 
as $\Delta m^{2}/E$.  This case is known as the MSW effect and was 
pointed out by Wolfenstein, Mikehyev, and Smirnov as being potentially 
important.  The quantum mechanical case can be found in many text books.  
To treat this effect with quantum field theory we have to calculate the 
Green's function for the neutrinos in matter.  The equation of motion 
derived from ${\cal L}^{\rm matter}_{\rm eff}(x)$ is in component form
\begin{equation}
(i \sslash{\partial} - m_{j})\nu_{j}(x) + \sum_{k}{\Aslash}_{jk}\nu_{k}(x) = \chi_{j}(x),
	\label{matter1}
\end{equation}
where $\chi_{j}(x)$ is the source.  Here the term responsible for the 
interaction with matter is given by contributions from neutral (N) and 
charge changing (C) currents in the static limit as

\begin{equation}
	A^{\mu}(x) = - \delta^{\mu 0}(A_{N}\id +A_{C}K), 
	\label{ matter1.1}
\end{equation}
where $K_{jk}=U^{*}_{ej}U_{ek}$ is the projector onto the electron neutrino 
flavor space.  The terms $A_{N}$ and $A_{C}$ have the values $A_{N} = 
-\sqrt{2}G_{F}N_{n}/2$, and $A_{C}=\sqrt{2}G_{F}N_{e}$ expressed in terms 
of the number density of neutrons or electrons, respectively.  For antineutrinos 
there is a change of sign in these terms.  

By transforming to momentum space and assuming that the 
neutrinos all have the same three-momentum $\bf p$, the Dirac equation in 
the presence of matter can be written in momentum space as

\begin{equation}
	(H\gamma^{0} -{\bf{p}}\cdot {\boldsymbol{\gamma}} \id - M  - (A_{N}\id + A_{C}K) \gamma^{0})u({\bf p})=0.
	\label{matter1.2}
\end{equation}

In this section, we will use the symbols $u_{M}$ for the solutions to the Dirac equation in the presence of matter.

The Eq. (\ref{matter1.2}) can be written as
\begin{equation}
 	H u_{M}({\bf p})=(-{\bf p} \cdot {\boldsymbol{\gamma}}\gamma^{0} + M\gamma^{0} + A_{N}\id + A_{C}K)u_{M}({\bf p})
 	\label{matter1.3}
\end{equation} 
To obtain the full set of roots one has, in the energy representation, to 
enlarge the flavor space with the two-dimensional energy space, by using 
$u_{\pm}=\frac{1}{2}(1\pm \gamma^{0})u$ for each flavor $f$.  This gives a 
$2f-$dimensional representation of the Hamiltonian, that will have $2f$ 
roots, i.e.  $f$ roots of both signs.  These roots are calculated for $f=2$ 
later.

It is then possible to diagonalize the Hamiltonian $H$, which is symmetric 
in this $2f-$dimensional representation, with a unitary transformation 
$U_{M}$ which is analogous to that one given e.g.  in Ref.  \cite{bib:ohls2000} 
for the case of an ordinary quantum mechanical description.  The Eq.~(\ref{matter1.3}) 
then becomes

\begin{equation}
	Hu_{Mj}= E_{j}u_{Mj},
	\label{matter1.4}
\end{equation}
where $u_{M}$ is related to $u$ in the mass representation by $u_{M}= 
U_{M}U^{\dagger}u$.  Since the flavor states $u_{f}$ are related to the 
matter states by $u_{f}=U^{\dagger}_{M}u_{M}$, where $U$ is naturally 
extended to the same dimension, the flavor states can be 
expanded in the mass representation also in case of neutrinos propagating 
through matter (of constant density).  For the antineutrinos a corresponding
equation holds with different sign on the matter terms, and the same transformation $U_{M}$ 
will diagonalize the Hamiltonian also for the antineutrinos.

The Feynman propagator, as a Green's 
function, is invariant under unitary transformations, and can therefore 
alternatively be expressed in the mass representation.

To obtain the transition amplitude we now utilize the same approach as in 
the previous section.  The neutrino field is expanded in the solutions to 
Eq.~(\ref{matter1.2}) and the spatial momentum plane waves.  The 
corresponding annihilation operator satisfies the equation

\begin{equation}
		a^s_j({\bf p},t) = a^s_j({\bf p}) e^{-iE_{j}t},
	\label{matter2.1}
\end{equation}
where $E_{j}$ now is the positive eigenvalue of $H$ and the two negative 
eigenvalues go with the adjoint creation operators. For the antiparticle 
operators these change their roles and the sign of the matter terms also 
change.

The amplitude of interest is in this case given by the expression
\begin{eqnarray}
	A^{rs}_{\alpha\beta \;{\rm matter}}&=&
	\int \frac{d^{3}q}{(2\pi)^{3}}\int \frac{d^{3}p}{(2\pi)^{3}} \frac{F_{\beta}^{\ast}({\bf X}_{d}, {\bf P}_{d}, {\bf q})}
	{\sqrt{2E_{\beta}({\bf q})}}\frac{F_{\alpha}({\bf X}_{s}, {\bf P}_{s},{\bf p})}
	{\sqrt{2E_{\alpha}({\bf p})}} {M}^{rs}_{M \alpha \beta}({\bf q}, {\bf p}) 
	,
	\label{matter4}
\end{eqnarray}
where 
\begin{eqnarray}
	{M}^{rs}_{M \alpha \beta}({\bf q}, {\bf p})& =&\sum_{jk}U_{M\beta j}U^{*}_{M\alpha k}{M}^{rs}_{M jk}({\bf q},{ \bf p}) \nonumber \\
	&= & \sum_{jk}U_{M\beta j}U^{*}_{M\alpha k} \int d^{3}x\int d^{3}y e^{ i{\bf p} \cdot {\bf x}} e^{-i{\bf q} \cdot {\bf y}}R^{rs}_{Mjk}(y-x),
	\label{matter5}
\end{eqnarray}

and the expression $R_{Mjk}^{rs}(x)$ is given by 
\begin{equation}
	R_{Mjk}^{rs}(x)= \int \frac{d^{4}k}{(2\pi)^{4}}e^{-ikx}\bar {u}^{s}_{M j}
	(q)i\gamma^{0}\langle 0|T(\nu_j(y)\nu_j^{\dagger}(x))|0\rangle \gamma^{0}u^{r}_{M k}(p).
	\label{matter6}
\end{equation}
Using the transformations given above and the fact that the time-ordered 
Green's function can be expressed in any basis, and thus can be replaced by 
the Feynman propagator in the mass representation, we arrive at the expression
\begin{equation}
		{M}^{rs}_{M \alpha \beta}({\bf q}, {\bf p}) =\sum_{jk}U_{\beta j}U^{*}_{\alpha k}
		 \int d^{3}x\int d^{3}y e^{i{\bf p} \cdot {\bf x}} e^{-i{\bf q} \cdot {\bf y}}{R}^{rs}_{jk}(y-x),
	\label{matter66}
\end{equation}
where
\begin{equation}
	{R}_{jk}^{rs}(x)= \int \frac{d^{4}k}{(2\pi)^{4}}e^{-ikx}\bar {u}^{s}_{j}(q)i\gamma^{0}S_{jk}(A,k)\gamma^{0}u^{r}_{k}(p).
	\label{matter6.1}
\end{equation}
The expression ${M}^{rs}_{M\alpha \beta}$ in Eq. (\ref {matter66}) will be our main task to calculate.

We first want to calculate the Feynman propagator in the presence of matter. What we seek is thus
the inverse $S(A,k)$ of the operator
\begin{equation}
  S^{-1}(A,k) =	G\gamma^{0}-{\bf k} \cdot {\boldsymbol{\gamma}} \id- M,
	\label{matter2}
\end{equation}
where $G=k^{0}\id - A_{N}\id - A_{C}K$ which operates in the tensor product space of the four 
dimensional Dirac space with the three dimensional mass space in the mass 
representation.

The neutral currents give a constant contribution to the energy that can be 
absorbed in its definition until the end of the calculation and will 
therefore be neglected from now on.  The charge current contribution, on the 
other hand, gives a more complicated contribution and will be dealt with 
explicitly.  For convenience we will use the notation $A\equiv A_{C}$ 
below.  The calculation can then readily be done and one finds

\begin{equation}
	S(A,k) = \left[G\gamma^{0}-{\bf k }\cdot {\boldsymbol{\gamma}} \id+ M +	
	(G\gamma^{0}-{\bf k} \cdot {\boldsymbol{\gamma}} \id + M)\gamma^{0}N^{-1} C \right] D^{-1},
	\label{matter3}
\end{equation}
where $N=G^{2}-{\bf k}^{2}\id -M^{2}$, $C= - A\left[ M,K \right]$, and 
$D=N-C N^{-1} C$.  This means that we are interested in the poles of 
$D^{-1}$ with respect to $k^{0}$.  We will perform the analysis for the 
case of two flavors below, leaving the three-flavor case for the future.

\subsection{The case of two flavors}
Here we will confine ourselves to calculate the amplitude explicitly only 
for two flavors, since we can then easily compare our result with the 
quantum mechanical expression for the resonance formula.

For two flavors we thus want to calculate the appearance amplitude
\begin{eqnarray}
	A^{rs}_{e\mu \;{\rm matter}}&=&
	\int \frac{d^{3}q}{(2\pi)^{3}}\int \frac{d^{3}p}{(2\pi)^{3}} \frac{F_{\mu}^{\ast}({\bf X}_{d}, {\bf P}_{d},{\bf q})}
	{\sqrt{2E_{\mu}({\bf q})}}  \frac{F_{e}({\bf X}_{s}, {\bf P}_{s}, {\bf p})}{\sqrt{2E_{e}({\bf p})}}{M}^{rs}_{M e\mu}({\bf q}, {\bf p}).
	\label{matter 6a}
\end{eqnarray}

Carrying out the integrations as before we end up with the expression
\begin{eqnarray}
		A^{rs}_{e\mu \;{\rm matter}}&=&
		\int \frac{d^{3}p}{(2\pi)^{3}} \frac{F_{\mu}^{\ast}({\bf X}_{d}, {\bf P}_{d}, {\bf p})}
	{\sqrt{2E_{\mu}({\bf p})}}\frac{F_{e}({\bf X}_{s}, {\bf P}_{s}, {\bf p})}{\sqrt{2E_{e}({\bf p})}}\nonumber\\
	&&\times \; i\int \frac{dk^{0}}{2\pi}e^{-ik^{0}(T_{d}-T_{s})}\bar {u}^{s}_{\mu}({\bf p})\gamma^{0}S(A,(k^{0},{\bf p}))\gamma^{0}u^{r}_{e}({\bf p}).
	\label{matter 6c}
\end{eqnarray}
We are thus interested in the expression
\begin{equation}
	{M}^{rs}_{M e\mu}=i\int \frac{dk^{0}}{2\pi}e^{-ik^{0}T}
	\bar {u}^{s}_{\mu}({\bf p})\gamma^{0}S(A,(k^{0},{\bf p}))\gamma^{0}u^{r}_{e}({\bf p}),
	\label{matter 6d}
\end{equation}
where $u_{\ell}=\sum_{j=1}^{2}U^{*}_{\ell j}u_{j}$, $\ell = e,\mu$, and $T=T_{d}-T_{s}$.

In the case of vacuum propagation, we have $M^{rs}_{e\mu} = \delta^{rs}\sum_{j} U^{*}_{e j}U_{\mu j}2E_{j}({\bf p})e^{-iE_{j}T}$, which in the high
energy limit goes to $\delta^{rs} e^{-iE T}2E \sin 2\theta \sin [T(E_{2}-E_{1})/2]$, with $2E = E_{1}+E_{2}$. In what follows, we will frequently omit the spin indices.

Let us first calulate the matrix $D = N - CN^{-1}C$ for the two flavors $e$ and $\mu$. The 
calculation is done in the mass eigenstate basis. 
In this case, we have
\begin{equation}
	G=k^{0}\id_{2} - AK.
	\label{2flavor1}
\end{equation}
Remember here that $k^{0}$ is really $k^{0} - A_{N}$.
Then, 
\begin{equation}
	C = - A\left[ M,K\right]= A(m_{2}-m_{1}) \left( 
	\begin{array}{cc}
		0 & K_{12}  \\
		-K_{21} & 0
	\end{array}
	   \right),
	\label{2flavor2}
\end{equation}
where $K_{21}=K_{12}= \sin \theta \cos \theta.$
Thus,
\begin{equation}
	C= \kappa \left( 
	\begin{array}{cc}
		0 & 1  \\
		-1 & 0
	\end{array}
	   \right)
	\label{2flavor3}
\end{equation}
with $\kappa= A(m_{2}-m_{1})K_{12}.$ We can then readily calculate the expression $CN^{-1}C$. 
The result is
\begin{equation}
	CN^{-1}C = -\frac{\kappa^{2}}{\det N}N^{T} =  -\frac{\kappa^{2}}{\det N}N,
	\label{2flavor4}
\end{equation}
since $N$ is symmetric. Thus, 
\begin{equation}
	D=N+\frac{\kappa^{2}}{\det N}N = \left( 1+\frac{\kappa^{2}}{\det N}\right) N
	\label{2flavor5}
\end{equation}
and
\begin{equation}
	D^{-1}= \frac{1}{\det N + \kappa^{2}}\left( 
	\begin{array}{cc}
		N_{22} & -N_{12}  \\
		-N_{12} & N_{11}
	\end{array}
	   \right).
	\label{2flavor6}
\end{equation}

In calculating the expression $\bar u \gamma^{0}S(A, (k^{0},{\bf p})) \gamma^{0}u$, we 
can use the fact that the spinors $\bar u$ fulfill the relations
\begin{equation}
	\bar u({\bf p}) ((E-AK)\gamma^{0}-{\bf p} \cdot {\boldsymbol{\gamma}} \id - M) = 0.
	\label{matter 7}
\end{equation}
This leads to two different terms
\begin{equation}
	S^{(1)} =  (k^{0}+E-2AK)D^{-1}
	\label{matter 8}
\end{equation}
and 
\begin{equation}
	S^{(2)} =  (k^{0}+E-2AK)N^{-1}CD^{-1}
	\label{matter 9}
\end{equation}
with
\begin{equation}
	\bar u_{i} \gamma^{0}S(A,(k^{0},{\bf p}))_{ij} \gamma^{0}u_{j} = u_{i}^{\dagger}S^{(1)}_{ij}u_{j} + \bar u_{i}S^{(2)}_{ij}u_{j}.
	\label{matter 9a}
\end{equation}
We are looking for the singularities of $D^{-1}$ and $N^{-1}CD^{-1}$.
The singularities of $D^{-1}$ are given by the roots with respect to 
$k^{0}$ of the equation
\begin{equation}
	N_{11}N_{22}-N_{12}^{2}+ \kappa^{2} = 0.
	\label{matter10}
\end{equation}
Now, $N$ can be calculated to be
\begin{equation}
	N= \left( 
	\begin{array}{cc}
    (k^{0})^{2}-{\bf p}^{2} -m_{1}^{2} + (A^{2}-2k^{0}A)K_{11} &  (A^{2}-2k^{0}A)K_{12}  \\
	(A^{2}-2k^{0}A)K_{12}	&  (k^{0})^{2}-{\bf p}^{2} -m_{2}^{2} + (A^{2}-2k^{0}A)K_{22}
	\end{array}
	   \right).
	\label{matter11}
\end{equation}
The equation (\ref{matter10}) for $k^{0}$ is then
\begin{eqnarray}
	((k^{0})^{2}-{\bf p}^{2} -m_{1}^{2} + 
	(A^{2}-2k^{0}A)K_{11})((k^{0})^{2}-{\bf p}^{2} -m_{2}^{2} + (A^{2}-2k^{0}A)K_{22})&-& \nonumber\\
	(A^{2}-2k^{0}A)^{2}K_{12}^{2}+\kappa^{2}=&0&.
	\label{matter12}
\end{eqnarray}

The other singularities are determined by the expression $N^{-1}CD^{-1}$. This can easily be calculated to be
\begin{equation}
	N^{-1}CD^{-1}= \frac{\kappa}{\det N +\kappa^{2}}\left( 
	\begin{array}{cc}
		0 & 1  \\
		-1 & 0
	\end{array}
	   \right).
	\label{matter13}
\end{equation}
Obviously, the singularities are again at the same location as the roots of Eq. (\ref{matter12}).

To solve Eq.~(\ref{matter12}) we first simplify the equation using the properties $K_{11}K_{22}-K_{12}^{2}=0$ 
and $K_{11}+K_{22}=1$, since $K$ is a projector.
We also use $E_{i}^{2}={\bf p}^{2}+m_{i}^{2} \;{\rm for}\; i=1,2.$ Then
\begin{equation}
	((k^{0})^{2}-E_{1}^{2})((k^{0})^{2}-E_{2}^{2})+A(A-2k^{0}) 
	((k^{0})^{2}-K_{11}E_{2}^{2}-K_{22}E_{1}^{2})+\kappa^{2}=0
	\label{matter14}
\end{equation}

Finally, we also have $\kappa^{2}=A^{2}(\Delta m)^{2}K_{11}K_{22}=A^{2}(\Delta m)^{2}\frac{1}{4}\sin^{2}2\theta$. 
The term $K_{11}E_{2}^{2} + K_{22}E_{1}^{2}$ can be simplified to be 
$\frac{1}{2}(E_{1}^{2}+E_{2}^{2}) +\frac{1}{2}(m_{2}^{2}-m_{1}^{2})\cos 2\theta.$
If we, for typographical reasons, put $k^{0}=x$ the equation can be written
\begin{eqnarray}
	x^{4}-2Ax^{3}+(A^{2}-E_{1}^{2}-E_{2}^{2})x^{2}+A(E_{1}^{2}+E_{2}^{2}
+(m_{2}^{2}-m_{1}^{2})\cos 2\theta)x & & \nonumber\\
-{\small \frac{1}{2}}A^{2}(E_{1}^{2} + E_{2}^{2}+(m_{2}^{2}-m_{1}^{2})\cos 2\theta) + 
{\small \frac{1}{4}}A^{2}(\Delta m)^{2}\sin^{2}2\theta && \nonumber \\
+ E_{1}^{2}E_{2}^{2}&= &0.
	\label{matter15}
\end{eqnarray}
We further introduce new variables $E=(E_{1}+E_{2})/2$ and $\Delta m^{2}=m_{2}^{2}-m_{1}^{2}$. Then,
 the equation above can be written
 \begin{eqnarray}
	x^{4}-2Ax^{3}+\left( A^{2}-2E^{2}-2\left(\frac{\Delta m^{2}}{4E}\right)^{2}\right) x^{2}+ A\left( 2E^{2}+ 2\left(\frac{\Delta m^{2}}{4E}\right)^{2}
+\Delta m^{2}\cos 2\theta \right) x & & \nonumber\\
-{\small \frac{1}{2}}A^{2} \left( 2E^{2}+ 2\left(\frac{\Delta m^{2}}{4E}\right)^{2} +\Delta m^{2}\cos 2\theta \right) + 
A^{2}\small \left({\frac{\Delta m^{2}}{2 (m_{1}+m_{2})}}\right)^{2}\sin^{2}2\theta & & \nonumber \\
+ E^{4}-2E^{2}\left(\frac{\Delta m^{2}}{4E}\right)^{2}+ \left(\frac{\Delta m^{2}}{4E}\right)^{4} = 0, & &
	\label{matter150}
\end{eqnarray}
which can be solved exactly. Let us call the roots of this equation $\lambda_{i}, \; i=1,\ldots,4.$

These roots are of course exactly the same as those we get by solving the 
secular equation for the Hamiltonian given in Eq.~(\ref{matter1.3}), when we 
enlarge the two-dimensional flavor space with the two-dimensional $u_{\pm}$ 
space related to diagonalizing $\gamma^{0}$, as was mentioned in 
Sec.~\ref{sec:matter1}.

Since we want to calculate the amplitude $M^{rs}_{e\mu}$ we are, in 
principle, interested in the expression
\begin{eqnarray}
	\bar u_{\mu}\gamma^{0}S \gamma^{0}u_{e}&=& -sc \bar u_{1}\gamma^{0}S_{11}\gamma^{0}u_{1} 
- s^{2} \bar u_{1}\gamma^{0}S_{12}\gamma^{0}u_{2}+ c^{2}\bar u_{2}\gamma^{0}S_{21}\gamma^{0}u_{1} +
cs \bar u_{2}\gamma^{0}S_{22}\gamma^{0}u_{2} \nonumber \\
&=&sc(\bar u_{2}\gamma^{0}S_{22}\gamma^{0}u_{2}-\bar u_{1}\gamma^{0}S_{11}\gamma^{0}u_{1}) 
+ (c^{2}\bar u_{2}\gamma^{0}S_{21}\gamma^{0}u_{1}- s^{2}\bar u_{1}\gamma^{0}S_{12}\gamma^{0}u_{2}),
	\label{matter15a}
\end{eqnarray} 
where $s \equiv \sin \theta$ and $c \equiv \cos \theta$.

The amplitude can be written in the form
\begin{equation}
M^{rs}_{e\mu}=	\bar u^{s}_{\mu}\gamma^{0}S\gamma^{0}u^{r}_{e}= \delta^{rs}\frac{1}{\det N +
\kappa^{2}}((k^{0})^{3}a_{3} + (k^{0})^{2}a_{2}+k^{0}a_{1} +a_{0}),
	\label{matter28}
\end{equation}
where the coefficients $a_{i}$ have to be calculated. We will discuss this point later and first 
calculate the amplitude.

\subsection{Calculation of the amplitude}
The  roots $\lambda_{i}$, $i=1,\ldots,4$ of Eq.~(\ref{matter150}) are all real, and  
we will proceed with the calculation of the amplitude. We will also assume that the roots are 
indexed so that in the limit $A\rightarrow 0$ the roots take on the values 
$\lambda_{1}=E_{1},\; \lambda_{2}=E_{2},\; \lambda_{3}=-E_{1}$, and $ \lambda_{4}=-E_{2}$.
We can then close the contour in the lower half-plane and calculate the residues of the amplitude 
avoiding the negative roots. The amplitude is then (minus)
the sum of the residues of the poles at $\lambda_{1}$ and $\lambda_{2}$.

The sum of the residues of the amplitude $M_{e\mu}$ is then
\begin{equation}
	M_{e\mu}= -i \frac{\lambda_{1}^{3}a_{3}+\lambda_{1}^{2}a_{2}+\lambda_{1}a_{1} +a_{0}}{(\lambda_{1}-\lambda_{2})
	(\lambda_{1}-\lambda_{3})(\lambda_{1}-\lambda_{4})} e^{-iT\lambda_{1}} -i \frac{\lambda_{2}^{3}a_{3}
	+\lambda_{2}^{2}a_{2}+\lambda_{2}a_{1}+a_{0}}{(\lambda_{2}-\lambda_{1})
	(\lambda_{2}-\lambda_{3})(\lambda_{2}-\lambda_{4})} e^{-iT\lambda_{2}}.
	\label{matter40}
\end{equation}
This expression is similar to that of the matrix element of the spectral resolution of
the evolution operator, only that here there is an influence from the negative energy terms through
the denominators. 

We first introduce the difference of the roots as $\Delta =\lambda_{2}-\lambda_{1}$.  
Next, we observe that we can 
translate all the roots by $A/2$ so that $x=y+A/2$ to make the sum of the new roots 
$y_{i}, i=1,\ldots,4$ satisfy $y_{1}+y_{2}+y_{3}+y_{4}=0.$ This transforms 
away the $x^{3}$-term in the equation.  Also the denominators in the 
amplitude are unaffected by this translation.  The numerators will obtain 
terms proportional to $A$ or $A^{2}$, which we can safely neglect.

In calculating the coefficients $a_{i}, \; i=0,\ldots,3$ we will neglect small 
terms in the numerator.  Our main interest is to locate the resonance 
region where the denominators are small.  We therefore first put the matter 
parameter $A$ to zero in the wave functions as well as in the terms $S^{(i)}, 
i=1,2.$ We further use the approximations $u^{s 
\dagger}_{i}u^{r}_{j}\approx \delta^{rs}(E_{i}+E_{j})$ for all $i,j$, 
and $\bar u^{s}_{i}u^{r}_{j}\approx \delta^{rs}(m_{i}+m_{j})$.  
For the diagonal elements the equality is strict.

In the limit of neglecting small terms in the numerator, we obtain the compact formulae:
\begin{eqnarray}
	a_{3}&\approx &\sin 2\theta \Delta m^{2}\frac{1}{2E},
	\label{matter29}\\
	a_{2}&\approx &\sin 2\theta \Delta m^{2},
	\label{matter30}\\
	a_{1}& \approx & \sin 2\theta \Delta m^{2} \frac{E}{2},
	\label{matter31}
\end{eqnarray}
and
\begin{equation}
	a_{0}=0.
	\label{matter31.1}
\end{equation}
The large contributions to the $a_{i}$:s given above all come from the $S^{(1)}$ term.

We now realize that introducing the small shift in 
the energy by the terms $A_{N}$, as we did in the beginning, amounts only to 
add $A_{N}$ to $A$ in the roots of Eq.~($\ref{matter150}$).  This is still a very small term 
and does not change anything essential in the above approximation, 
especially not in the denominators, that contain differences of the roots.  
Apart from appearing in the small terms neglected, the contribution from 
$A_{N}$ appears in the overall phase factor as an addition to $A$.  
However, this will not affect the probability.

Finally, we approximate the sums and products of roots in terms of $y_{i}$ with 
their limiting values as $A=0$ (and $A_{N}=0)$. This leads to an 
expression for the amplitude of the form

\begin{eqnarray}
	M_{e\mu} &\approx  & -i \frac{1}{4E \Delta}e^{-iT(y_{1}+y_{2}-A)/2} \left\{ -i\left( a_{3}2E^{2}+2E a_{2}+ 2a_{1} +\frac{2a_{0}}{E}\right) \sin (T\Delta /2)  \right.   \nonumber\\
	&& \left.   + \left( a_{3} \Delta m^{2}+\frac{ a_{2}\Delta m^{2}}{2E} 
	 - \frac{a_{0}\Delta m^{2}}{2E^{3}} \right) \cos(T\Delta /2) \right\}.
	\label{matter41}
\end{eqnarray}
Inserting the values for the coefficients $a_{i}$ from Eqs.~$(\ref{matter29}) - (\ref{matter31.1})$, we 
obtain the formula
\begin{equation}
	M^{rs}_{e\mu}\approx -i \delta^{rs}e^{-i(E-A/2)T} 2E \sin 2\theta_{M}\left( \frac{\Delta m^{2}}{4E^{2}}\cos (T\Delta/2)-i\sin (T\Delta/2)\right),
	\label{matter42}
\end{equation}
where 
\begin{equation}
	\sin 2\theta_{M} = \frac{\Delta m^{2}\sin 2\theta}{2E \Delta}.
	\label{matter43}
\end{equation}
In the limit $A=0$, we have $\Delta=\Delta m^{2}/2E$. Neglecting the 
term of order $\Delta m^{2}/E^{2}$, we obtain
\begin{equation}
	M_{e\mu}(A=0)= e^{-iET}2E\sin 2\theta \sin(T\Delta m^{2}/4E),
	\label{matter44}
\end{equation}
which is the previously derived result without matter effects and coincides with the quantum mechanical expression.  

We can simplify this result in Eq.~(\ref{matter42}) further by observing that 
$\Delta = (y_{2}^{2}-y_{1}^{2})/2E$ within the approximation scheme 
we are envisaging. We will use a perturbative approach to obtain approximate 
analytic expressions for the roots $y_{i}$. First, we omit the linear term in Eq.~(\ref{matter150}) after 
translating to $ x=y+A/2$ and solve the resulting equation in $y$ exactly. We then use dimensional analysis to make the {\it Ansatz\/} 
\begin{eqnarray}	
	y_{1}(\epsilon)& =& \sqrt{\frac{1}{4}A^{2}+ B^{2}-\sqrt{A^{2}B^{2} - A^{2}S^{2} +4a^{2} - A\Delta m^{2}\cos 2\theta \epsilon}}, \\
	y_{2}(\epsilon)&=& \sqrt{\frac{1}{4}A^{2}+ B^{2}+\sqrt{A^{2}B^{2} - A^{2}S^{2} +4a^{2} - A\Delta m^{2}\cos 2\theta \epsilon}},
	\label{roots100}
\end{eqnarray}
where $B^{2}=E^{2}+(\frac{\Delta m^{2}}{4E})^{2}$ and $a=\Delta m^{2}/4$. Inserting this into Eq.~(\ref{matter150}) 
gives the equations
\begin{equation}
	\epsilon=y_{i}(\epsilon),\; i=1,2.
	\label{roots101}
\end{equation}
Solving to lowest order gives $\epsilon_{1}=\sqrt{B^{2}-2a} =E_{1}$ and $\epsilon_{2}=\sqrt{B^{2}+2a}=E_{2}$.
The difference of the roots is then
\begin{eqnarray}
	\Delta & \approx & \frac{1}{2E}\left( \sqrt{A^{2}B^{2} - A^{2}S^{2} +4a^{2} - A\Delta m^{2}\cos 2\theta \sqrt{B^{2}+2a} }\right.\nonumber\\
	&& +\left. \sqrt{A^{2}B^{2} - A^{2}S^{2} +4a^{2} - A\Delta m^{2}\cos 2\theta \sqrt{B^{2}-2a}} \right).
	\label{roots102}
\end{eqnarray}
This gives
\begin{equation}
	\sin 2\theta_{M}= \frac {\Delta m^{2} \sin 2\theta}{2E\Delta}=\frac{2 \sin 2\theta}{\tilde \Delta},
	\label{roots104}
\end{equation}
where
\begin{eqnarray}
	{\tilde \Delta}& =& \sqrt{ (AB/2a)^{2} -{( AS/2a)}^{2}+1- (A/a)\cos 2\theta \sqrt{B^{2}+2a}} \nonumber \\
	&& + \sqrt{(AB/2a)^{2} -{ (AS/2a)}^{2}+1-(A/a)\cos 2\theta \sqrt{B^{2}-2a}}.
	\label{roots105}
\end{eqnarray}
The minimum of $\tilde \Delta$ is given up to small terms by
\begin{equation}
	\cos 2\theta = \frac{A}{4a}(\sqrt{B^{2}-2a}+\sqrt{B^{2}+2a}) = \frac{2EA}{\Delta m^{2}}.
	\label{roots103}
\end{equation}
This minimum location coincides with the quantum mechanical location. However, the 
minimum value is not exactly the same as in the quantum mechanical case due to 
small extra terms. One of these terms is dependent upon the sum of the neutrino 
masses rather than their mass squared differences. 

The expression for $\tilde \Delta$ can be rewritten as
\begin{eqnarray}
	{\tilde \Delta}& = &\sqrt{ (A\sqrt{B^{2}+2a}/2a -\cos 2\theta)^{2} +\sin^{2}2\theta(1- A^{2}/(m_{1}+m_{2})^{2}) - A^{2}/2a} \nonumber \\
	&&+\sqrt{(A\sqrt{B^{2}-2a}/2a -\cos 2\theta)^{2} +\sin^{2}2\theta (1-A^{2}/(m_{1}+m_{2})^{2}) + A^{2}/2a}.
	\label{roots105.1}
\end{eqnarray}
This is the resonance formula. The minimum at the resonance peak
is dependent on the neutrino mass in the very small terms:
\begin{equation}
	{\tilde \Delta}_{\rm peak} =2\sin 2\theta \left( 1- \frac{1}{2} \frac{A^{2}}{(m_{1}+m_{2})^{2}} +\ldots \right).
	\label{delta3}
\end{equation}

With realistic values of $A \approx 10^{-12}$ eV, $a\approx 10^{-5}$ e${\rm 
V}^{2}$, and $B$ in the MeV range or higher, consistent with the 
experimental mass differences and the possible matter densities, the
corrections are of the order of $10^{-14}$, and the value 
of $\tilde \Delta$ at the minimum is given by ${\tilde \Delta} 
=2\sqrt{\sin^{2} 2\theta (1+{\cal O}(A^{2}/(m_{1}+m_{2})^{2}) +{\cal 
O}(A^{2}/\Delta m^{2})} \approx 2\sin 2\theta $ as in the quantum 
mechanical case.  This indicates that the other corrections would be still 
smaller.  The full expression for $\Delta$ can be calculated using the 
formulas above without approximations, but will be quite difficult to 
handle and is not of much use.  Our expression Eq.~(\ref{matter42}) contains 
the field theoretic roots to the 
equation.  These roots are dependent not only on the mass squared 
differences, but also weakly dependent on the sum of the neutrino masses.  
This comes from the term $\kappa$ in the equation for the singularities.  
This term appears from the nontrivial coupling of the two mass eigenstates 
in the calculation of the inverse propagator.

\section{Summary and conclusions}
In summary, we have discussed field theory methods to treat neutrino 
oscillation phenomena with wave packets for neutrino flavor states in 
vacuum and in matter with constant density.  We have carried out the 
calculation of the oscillation amplitude with matter effects for 
the case of two flavors explicitly.  The resulting formula coincides with the quantum 
mechanical expression using plane waves, up to very small terms.  These 
terms depend on the neutrino masses, but are unfortunately too small to be 
observable at the present time.  To the best of our knowledge this is the 
first detailed calculation of the neutrino oscillation formula using field 
theory and integrating out the poles in the Green's function of the 
neutrinos in the presence of matter.  Earlier works, e.g. in Ref. \cite{bib:Fujii,bib:Cardall2} 
have used various approximations at an earlier stage and thereby not being 
able to isolate the poles exactly.

Our calculation is done in what is essentially the rest frame of the 
detector and source, using wave packets, in which the neutrinos have the 
same three-momenta.  It should be possible, though more complicated, to 
relax this condition, and let also the three-momenta in the superposition 
vary within the limits allowed by the wave packet parameters.

For three flavors the corresponding equation for the poles will be of sixth 
order, and can therefore not in general be solved exactly.  We 
nevertheless believe that approximate analytic solutions to the equation can 
be found that can be used to study the singularity structure, in a way 
analogous to the two flavor case treated here.

\section*{Acknowledgment}

This work was carried out with financial support from the Swedish Research 
Council, contract 621-2001-1978 and G\"oran Gustafssons Stiftelse.  We 
gratefully acknowledge useful discussions with S. Bilenky, T. Ohlsson, and W. Winter.

\end{document}